\documentclass[journal,12pt,onecolumn,draftclsnofoot]{IEEEtran}
\usepackage{cite}
\usepackage{amsmath,amssymb,amsfonts}
\usepackage[font=footnotesize,caption=false,subrefformat=parens,labelformat=parens]{subfig}
\usepackage[bookmarks=false]{hyperref}
\usepackage{graphicx}
\usepackage{textcomp}
\usepackage[center]{caption}
\usepackage[normalem]{ulem}
\usepackage{makecell}    
\begin{document}
\pdfoutput=1
\bstctlcite{IEEEexample:BSTcontrol}
\title{
Multilayer Minkowski Reflectarray Antenna with Improved Phase Performance}

\author{
    \IEEEauthorblockN{Ender Ozturk\IEEEauthorrefmark{1}, Birsen Saka\IEEEauthorrefmark{2}}\\
    \IEEEauthorblockA{\IEEEauthorrefmark{1}NC State University, Raleigh, NC
    \\ E-mail: \textit{\{eozturk2\}}@ncsu.com}\\
    \IEEEauthorblockA{\IEEEauthorrefmark{2}Hacettepe University, Ankara, Turkey
    \\ E-mail: \{\textit{birsen}\}@ee.hacettepe.edu.tr}}

\makeatletter
\begin{center}
{\fontsize{18pt}{14pt}\selectfont\bfseries\@title\par}
\@author
\end{center}

\begin{abstract}
We propose a multi-layer unit cell consisting of Minkowski fractal shaped reflector with aperture coupled phasing stubs to obtain a broad phase range for the reflectarray antenna with smaller unit cell size and inter-element spacing compared to some other studies in the literature. In the study, simulations are conducted using HFSS™ program with Floquet Port incidence. A unit cell with two cycles of phase range and very low reflection loss is designed. 21.9\% shrinkage is achieved in patch surface area using Minkowski fractals. Subsequently, proposed unit cell is used to design and fabricate a full scale 221 element reflectarray antenna.  At 10 GHz, simulation results are compared with measured data and good agreement is observed. 
\end{abstract}

\begin{IEEEkeywords}
aperture coupling, arrays, fractal antennas, reflectarrays
\end{IEEEkeywords}

\section{Introduction}
\label{sec:introduction}
Reflectarrays are planar reflector antennas well-known for their various innate advantages. They are highly versatile, easy to design and fabricate since no feed circuit is needed. They also have low profile and reduced cost, volume, and weight. However, they are known to suffer from low efficiency and bandwidth~\cite{huang2007reflectarray}.

The pioneering work on reflectarrays started with the investigations of Berry et al. in 1963~\cite{Berry_reflectarray}. The first reflectarray consisted of a feed antenna and a set of rectangular waveguides. The length of the waveguides are arranged such that the field received from the front aperture is reflected back with an intended phase contribution to form a beam at a designated direction. Microstrip antennas substituted waveguides beginning from~\cite{Maligisi1978_microstrip} in reflectarray studies. Research on reflectarrays until the late 90s can be grouped by considering the microstrip reflector types and the phase variation mechanisms. 

Designs in the first group mainly use phase delay stubs. In this technique, incident field couples with the main reflector and travels to the end of a matched stub and reflects back. Subsequently, the signal is reradiated with a phase difference proportional to the length of the stub line\cite{Huang_microstrip_reflectarray}. In the second group of studies, microstrip dipoles are used as reflectors and phase variation is provided by changing the length of the dipoles~\cite{Kelkar_FLAPS}. Third group is characterized with designs that realize the phase variation by changing the resonance dimensions of the reflecting patches~\cite{Pozar_analysis_reflectarray}. Antennas in this group exhibit quite good performance for various parameters, such as cross polarization levels, since there are no bent stubs with respect to the first group. However, phase ranges occur less than a full
cycle for single layer designs.

One additional technique to provide phase variation is rotating the patches by some mechanism such as micro-motors~\cite{Huang_Ka_band, Huang_bandwidth_study}. This technique is extremely useful when working with circularly polarized waves.

In the last two decades, studies on reflectarrays showed a wide range of progress in different directions. Most research focused on improving beam steering capability, enhancing frequency bandwidth, reducing cross polarization levels and mutual coupling, developing new analysis techniques to better and/or faster predict the behavior of a unit cell in a full size antenna and miniaturizing antennas.

Multilevel reflectarray antennas are introduced to overcome the phase range problem of single layer reflectors with which the phase variation is provided by changing the resonance dimension of the patch. Two~\cite{Encinar_design_of_two_layer} and three~\cite{Encinar_three_layer} layer reflectarrays exhibit good performance in terms of phase range and smoothness as well as frequency bandwidth. Multilayer structure makes it also possible to realize phase variation by the placement of a stub below main reflector coupled via an aperture. Two of the first examples are~\cite{Bialkowski_passive_reflectarray} and~\cite{Keller}. In~\cite{riel_design_beam_scanning} and~\cite{venneri_design_validation}, varactor loaded microstrip lines are used to realize the phase variation. In\cite{Carrasco_aperture_coupled, Carrasco_slots}, Carrasco et al. achieved more than 4-cycle phase range with less than 0.8 dB reflection loss. The main advantages of this technique are listed as follows. First, since the stub is not directly attached to the patch, there is more room in the cell which can be utilized to enhance the phase range. Second, since an aperture is used between the stub and the main reflector, most of the energy could be held in the upper hemisphere, thus reflection loss stays low.  

In the context of antenna miniaturizing, fractal shaped reflectors became prominent recently. Infinitely iterated fractals have infinite circumference at a finite reach. This feature opens a new area of development for reflectarray antennas in terms of miniaturization.

Even though there exist some studies on Sierpinski~\cite{Borja_sierpinski} and Koch~\cite{Romeu_Koch} type fractals, Minkowski fractals are the ones that have been studied most in the literature. In Refs.~\cite{Zubir_minkowski, Costanzo_fractal, Costanzo_miniaturized}, single layer Minkowski fractals are examined. In these works, it is seen that maximum phase range provided by the unit cell designs are insufficient (lower than one cycle). In our previous study, we designed a single layer Minkowski fractal that exceeds one cycle~\cite{Ozturk_double_orthogonal}. In that study, we proposed a single layer Minkowski fractal shaped unit cell with coplanar double orthogonal stubs. Due to the existence of stubs, we were not able to fully exploit Minkowski fractal’s miniaturization advantage. In this study, however, the phasing stubs are moved below the main reflector. Larger surface size reduction, twice phase range, slightly better directive gain and much better cross polarization performance are achieved at the cost of a thicker antenna. 

Multilevel Minkowski fractal reflectarrays have not been investigated in detail. In~\cite{Wahid_dual_layer_minkowski}, only simulation results for a double layer Minkowski reflectarray unit cell are reported which has more than 450° phase range.  
\begin{figure}[!thbp]
\centerline{\includegraphics[width=0.7\columnwidth]{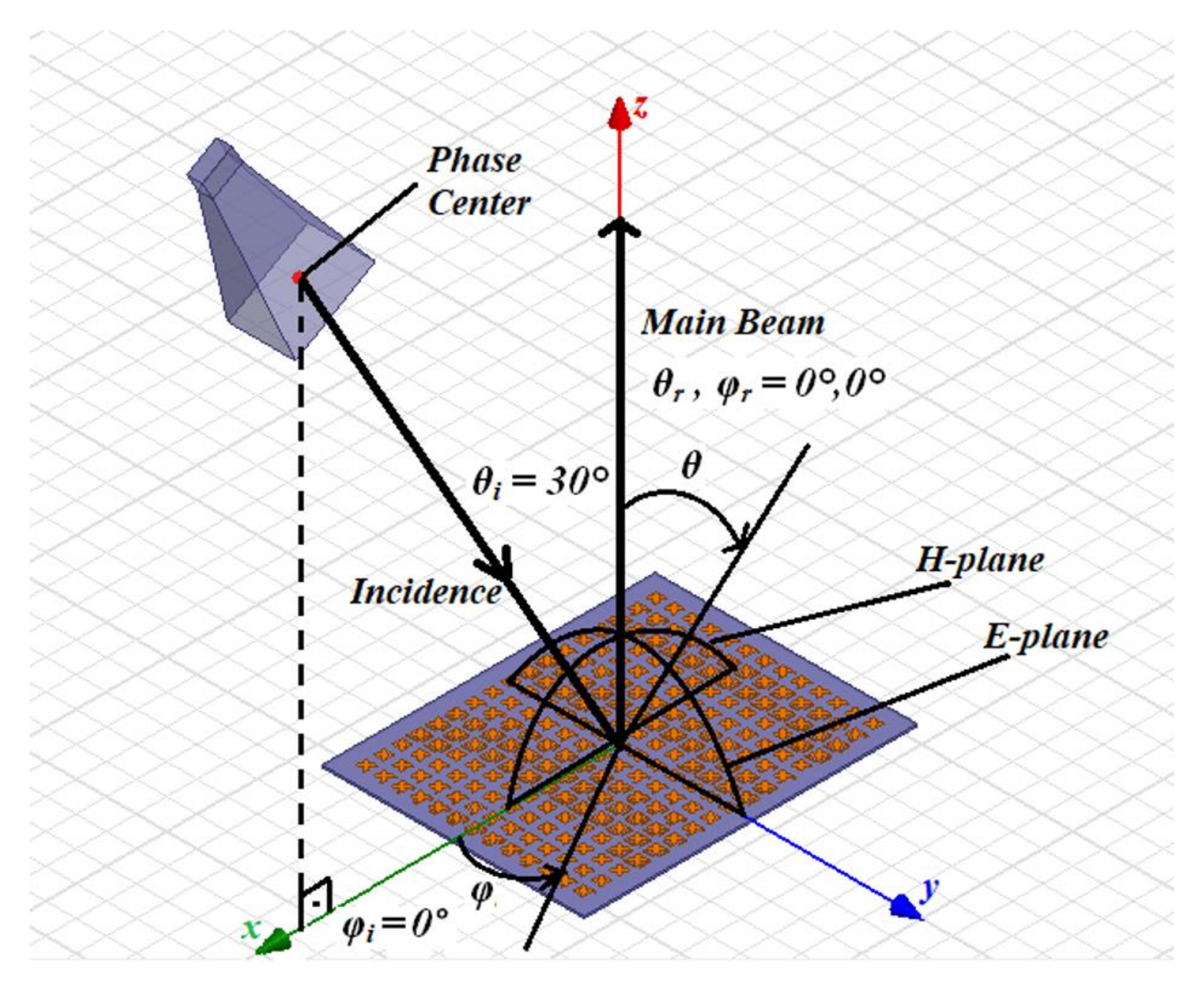}}
\caption{Reflectarray antenna geometry.} 

\label{Fig:reflectarray_geometry}
\vspace{-3mm}
\end{figure}
In this paper, in accordance with the study in~\cite{E_ozturk_phd}, a multilayer Minkowski fractal reflectarray with aperture coupled phasing stub is designed. Our main objective is to solve the insufficient phase range problem of Minkowski fractal reflectarrays which has been addressed in the literature. A 221 element full size antenna is fabricated and measured to show the performance of the proposed unit cell.

\section{Reflectarray Theory}

Reflectarrays are flat surfaces of cells with each having its own shape to contribute to reflected field in order to form a collimated beam (Fig.~\ref{Fig:reflectarray_geometry}). A unit cell is, in general, backed by ground planes, and includes single or multi-dielectric layers. Unit cells have main reflectors that are tuned to resonance frequency of the antenna. Reflected field from each cell can be written as a superposition of fields reflected from the ground plane and that scattered from the patch~\cite{Pozar_millimeter_wave}.

\begin{align}
    \begin{bmatrix}
        E_{\theta}^r\\
        E_{\phi}^r
    \end{bmatrix}
&= \left(
    \begin{bmatrix}
        R_{\theta \theta} & 0\\
        0 & R_{\phi \phi}^r
    \end{bmatrix}
    +
    \begin{bmatrix}
        S_{\theta \theta} & 0\\
        0 & S_{\phi \phi}^r
    \end{bmatrix}
\right)
\begin{bmatrix}
        E_{\theta}^i\\
        E_{\phi}^i
    \end{bmatrix}\\
& e^{jk_0(x sin\theta_i~cos\phi_i + ysin\theta_i~sin\phi_i - zcos\theta_i) \nonumber}
\end{align}

where $E_{\theta}^r$, $E_{\theta}^i$ and $E_{\phi}^r$, $E_{\phi}^i$ are reflected and incident field components at $\theta$ and $\phi$ polarizations. $R_{\theta \theta}$ and $R_{\phi \phi}$ are coefficients due to reflection from substrate (given in closed form in~\cite{Frances_phd}) and $S_{ab}$ are coefficients due to scattering from conducting patches. In order to design a full scale reflectarray, behavior of unit cell in the presence of neighboring cells should be calculated as accurately as possible. To do this, $S_{ab}$ coefficients can be calculated using periodic boundary conditions. This approach is formulated by the Floquet Theory. Each field component on the cell is written as the sum of an infinite number of modes, called Floquet modes, in a periodic environment, and solved with a numerical technique or software of choice.

\begin{figure*}[hbtp]
\centering
\captionsetup[subfigure]{labelformat=parens}
\setcounter{subfigure}{0}
\subfloat[]{\includegraphics[width=0.3\linewidth]{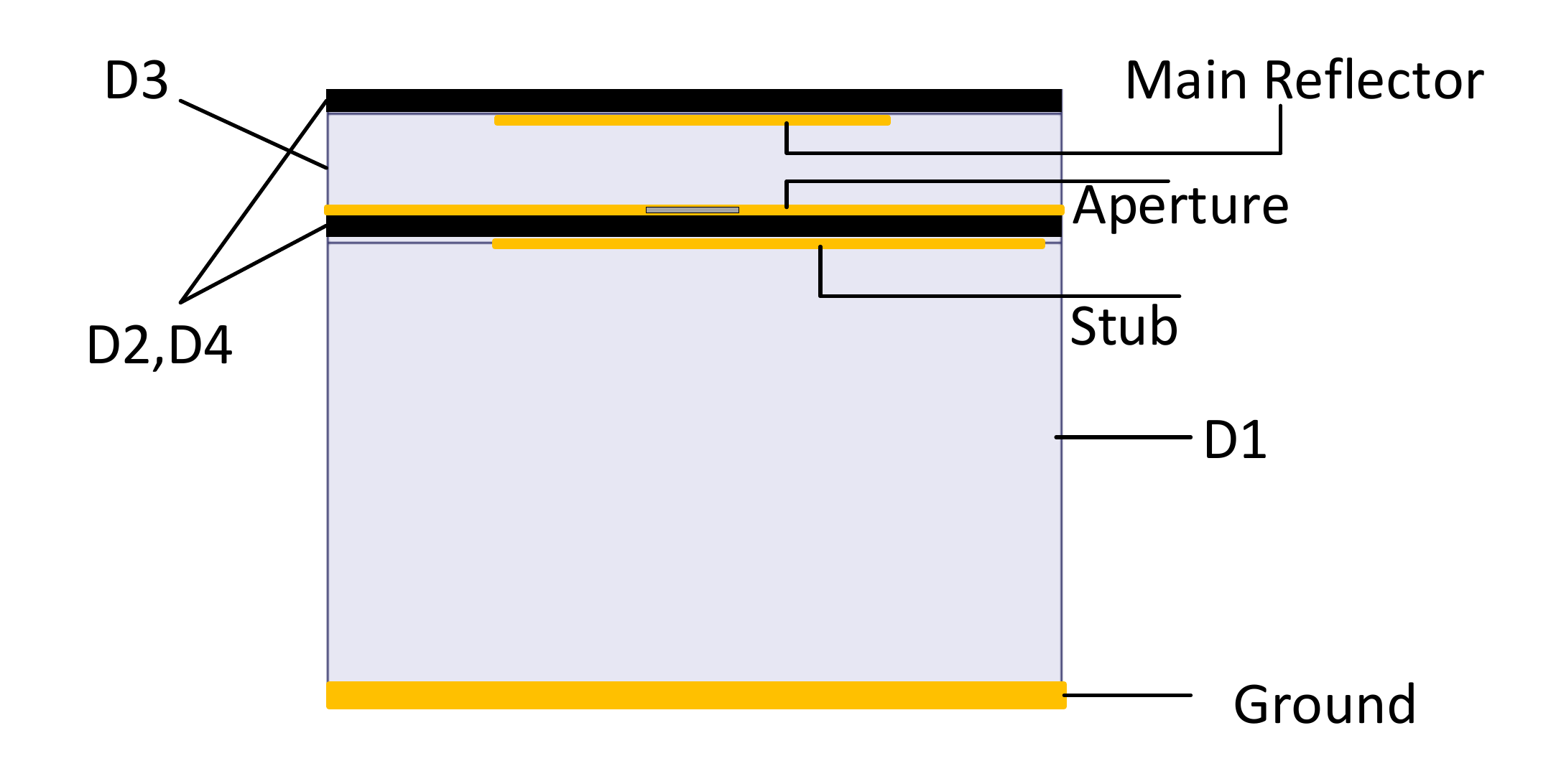}
\label{fig_noised_plots_a}}
\subfloat[]{\includegraphics[width=0.15\linewidth]{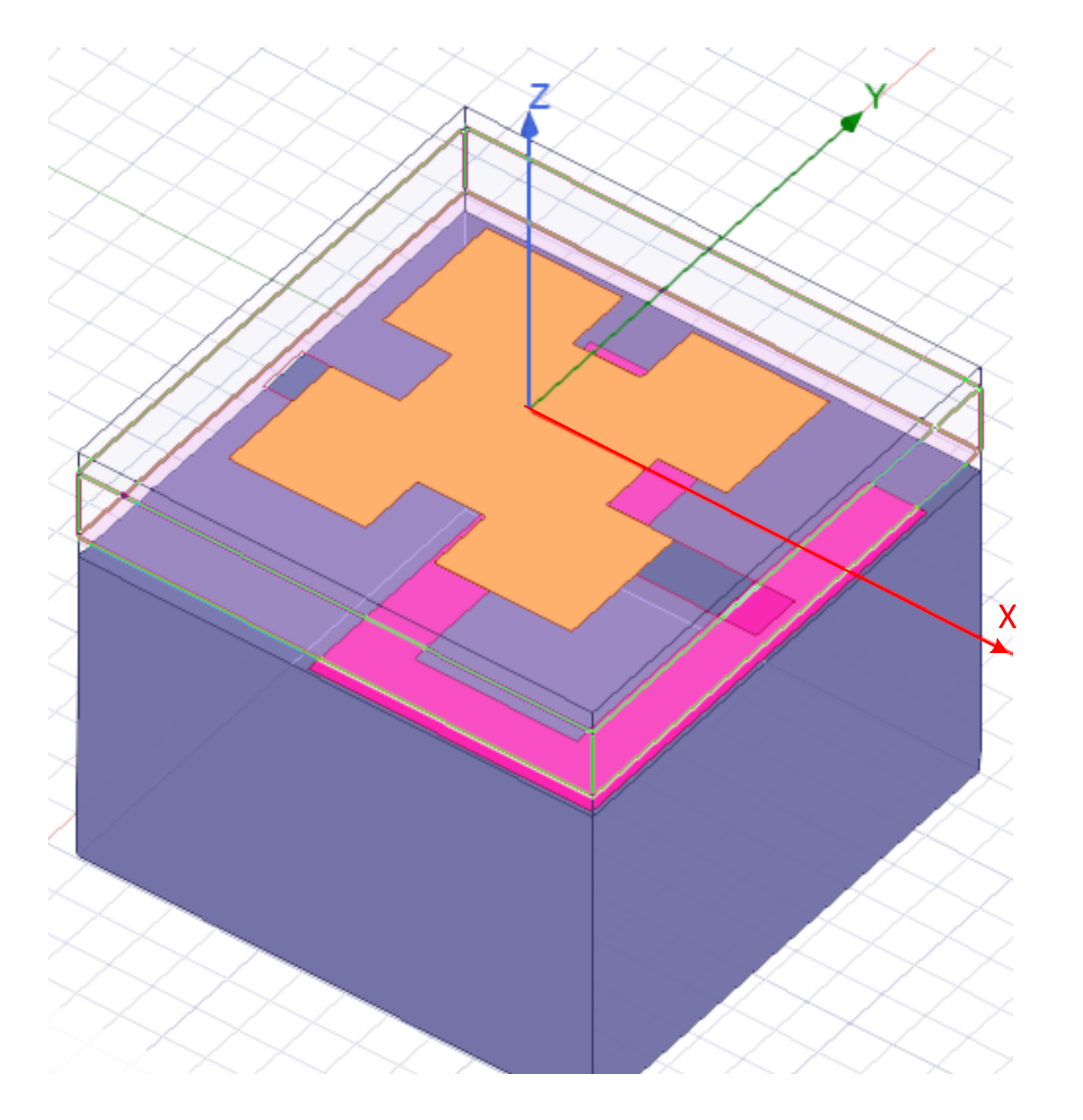}
\label{fig_noised_plots_b}}
\subfloat[]{\includegraphics[width=0.15\linewidth]{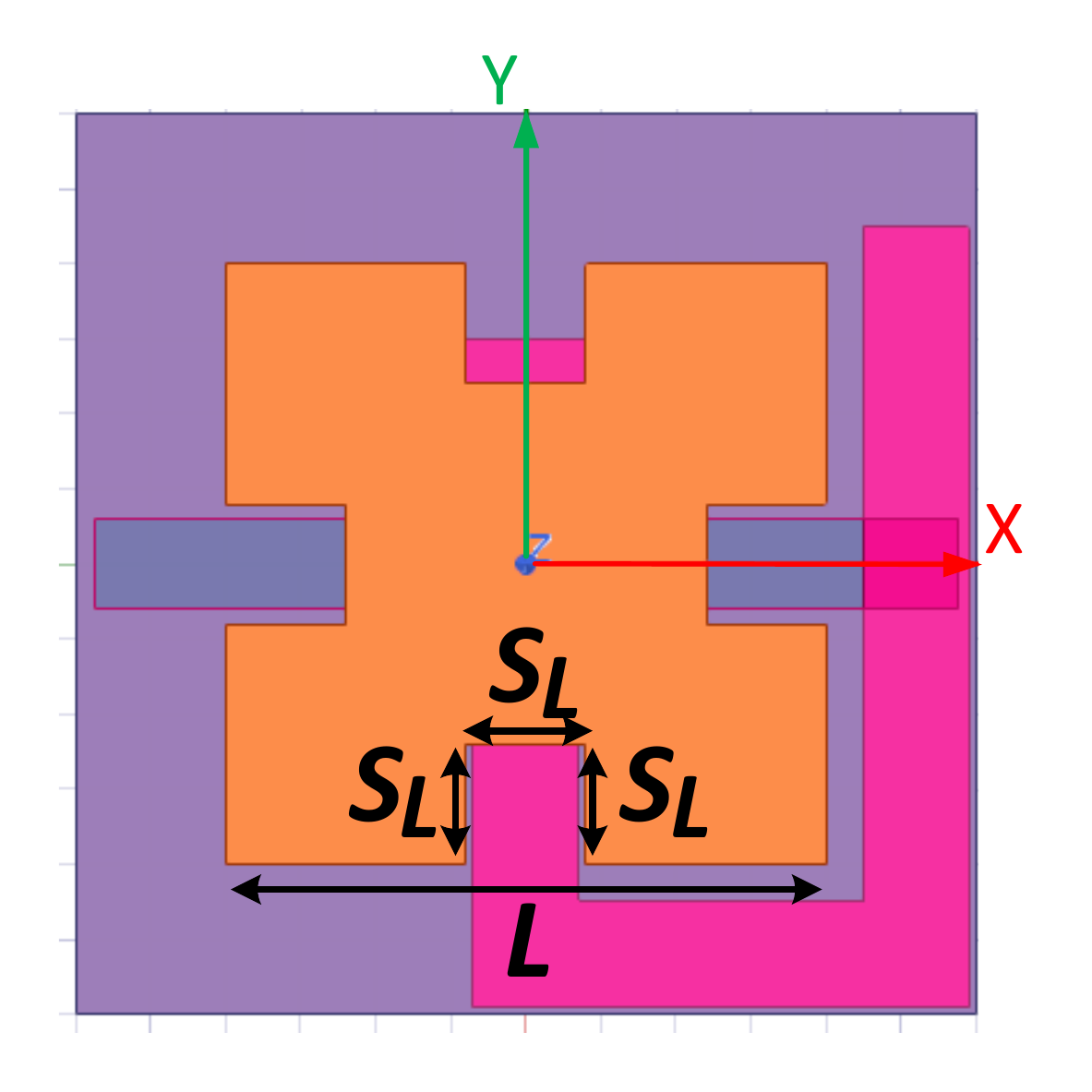}
\label{fig_noised_plots_c}}
\subfloat[]{\includegraphics[width=0.15\linewidth]{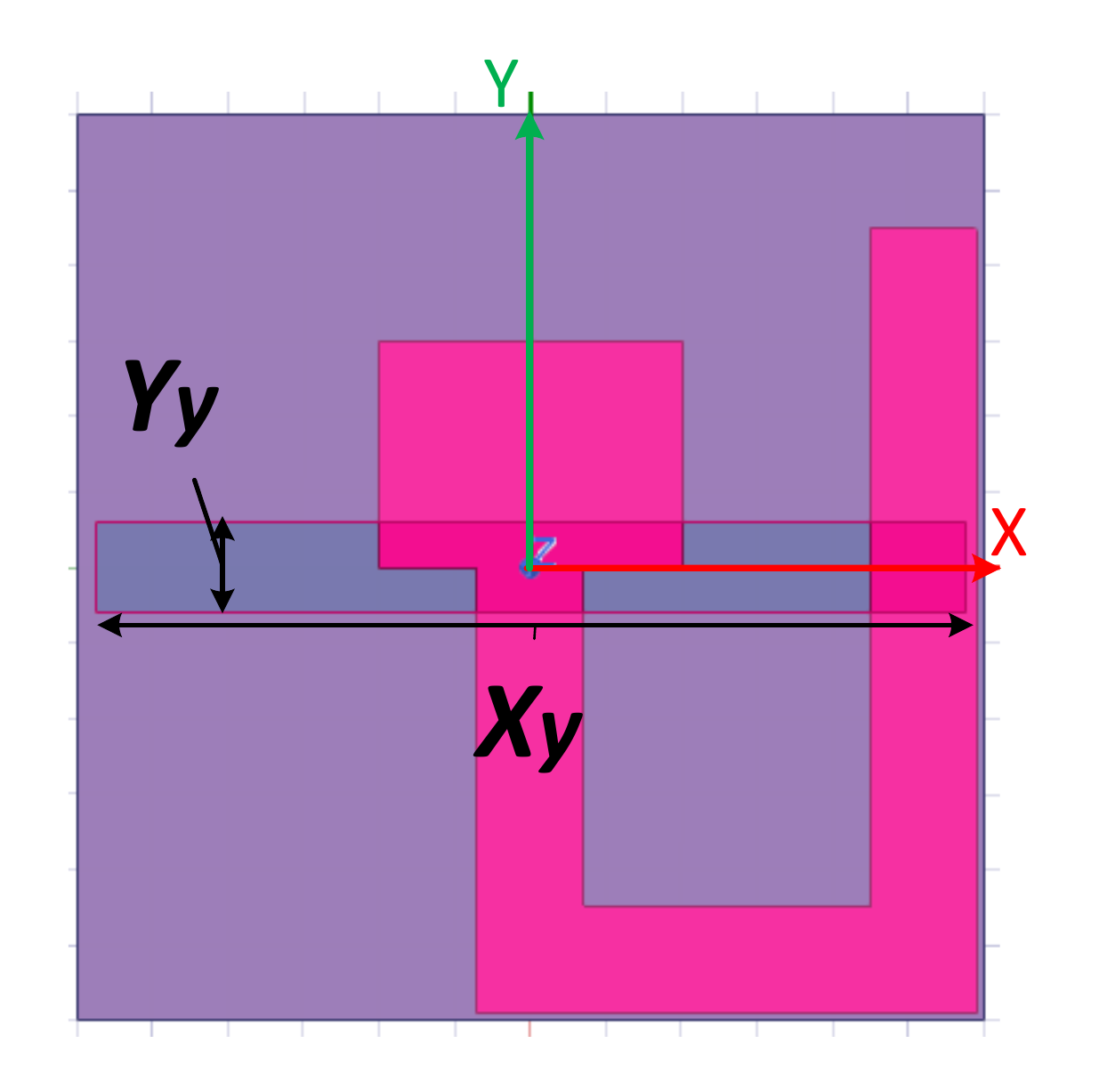}
\label{fig_noised_plots_d}}
\subfloat[]{\includegraphics[width=0.15\linewidth]{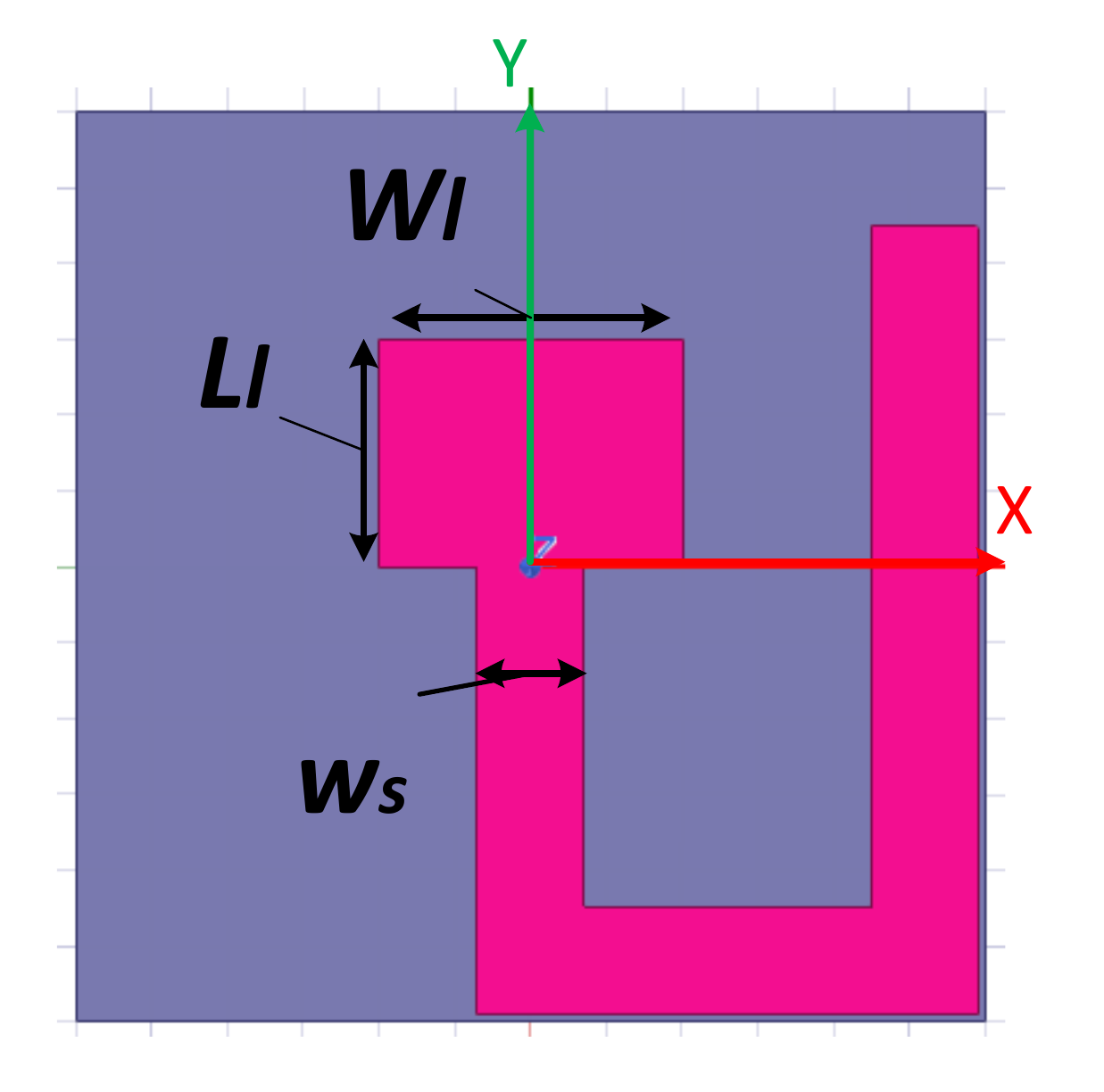}
\label{fig_noised_plots_e}}
\caption{Minkowski fractal multilayer unit cell geometry (a) Layers in profile, (b) Isometric view, (c) Bottom side of D2 dielectric layer, (d) Top side of D4 dielectric layer, (e) Bottom side of D4 dielectric layer. }
\label{Fig:unit_cell}
\vspace{-3mm}
\end{figure*}

\section{Unit Cell Design}
The main feature of a fractal structure is to increase the edge length without increasing the area. Therefore, it is possible to have smaller patches for the same resonance frequencies in comparison to the standard squares or rectangles. In our study, the unit cell of the designed reflectarray is a Minkowski square fractal. The proposed Minkowski fractal multilayer unit cell geometry is shown in Fig.~\ref{Fig:unit_cell}. In the structure, there are two dielectric layers separated by an air gap (D3). Main reflector is at the bottom side of upper layer (D2). Top side of lower layer (D4) is covered with copper cladding with hollowed apertures facing the main reflector. Centers of the aperture and the main reflector both lie on the z-axis.  There are two stubs at the bottom side of lower layer. The stub which is along the y-axis is the matching stub. The stub lying towards the opposite side is the phasing stub. Center of the connection point of these two stubs also lie on the z-axis. Reflection phase of the unit cell is arranged by changing the length of this stub. There is the ground plane at the bottom of the unit cell and another air gap between ground and bottom dielectric layer (D4). 

Phase variation is provided by a stub which is carried beneath the cell with an aperture in between making it possible for main reflector to couple to the stub. This scheme allows more space for the stub and greater phase range with the cost of increased thickness. Similar designs  were implemented for ordinary square reflectors in~\cite{Carrasco_aperture_coupled} and~\cite{Carrasco_slots}.

First iteration Minkowski fractal is formed by extracting an $SLxSL$ square from the middle of each edge of an $LxL$ square. Scale factor $S$ is chosen to be 0.2 in this study. 
\begin{figure}[!hbtp]
\centerline{\includegraphics[width=0.8\columnwidth, height = 19cm]{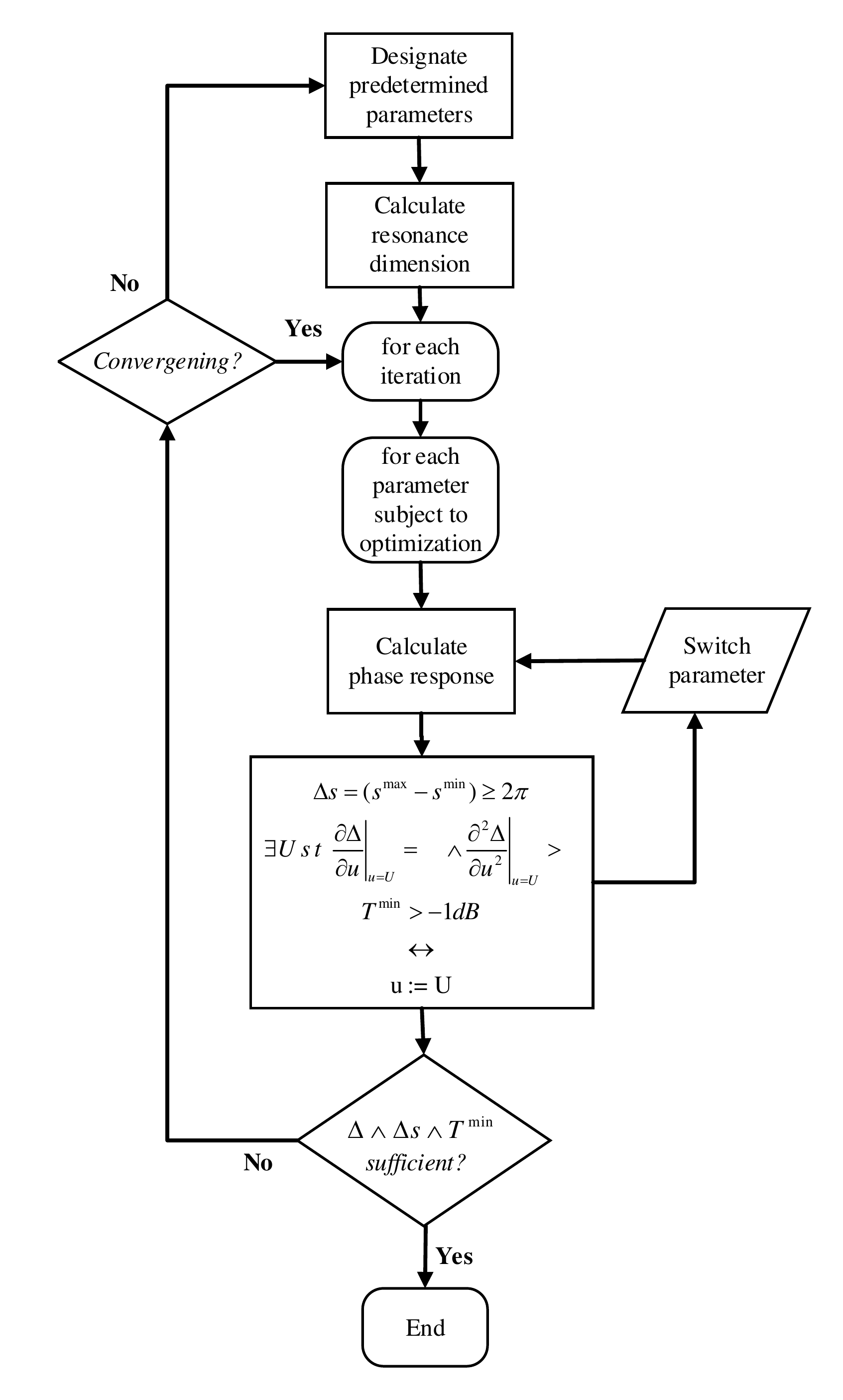}}
\caption{Optimization procedure flow chart. }
\label{Fig:optimization_flow}
\end{figure}
An optimization procedure is implemented to determine the dimensions of unit cell under certain constraints. Steps of the implemented algorithm are given in Fig.~\ref{Fig:optimization_flow}. $\Delta s$ is the maximum phase range, $T^{\text{min}}$ is the maximum return loss, $u$ is the parameter to be optimized, $U$ is the value of $u$ at the optimal point, and $\Delta$ stands for the error function which reflects proximity of phase curve to the desired phase line. Desired phase line is defined as the line having constant slope which begins and ends at the same points with the phase curve. $\Delta$ is calculated for $n$ samples using
\begin{align}
    \Delta = \sum_{k=0}^n \sqrt{|s_k^2 - {s_k^i}^2|}
\end{align}
where $s_k$ is the phase value for a parameter and $s_k^i$ is the corresponding point on the desired line. 

Ansys® HFSS™ (High Frequency Structure Simulator) with Floquet Port model is employed in this work. First, a unit cell model is created using values given in Table~\ref{tab:predetermined} for predetermined parameters, and nominal values for parameters subject to optimization. Resonance dimension of the patch is determined using the software. Subsequently, error function, phase range and maximum return loss are calculated for each parameter subject to optimization. When the return loss and phase range values are sufficient, the value of the parameter that minimizes error is designated as the new value for that parameter in the design. Then algorithm moves to the next parameter until all optimization parameters are updated, which finalizes the first iteration. If phase curve linearity, return loss and phase range values are insufficient, further iterations are needed. If no convergence is reached after several iterations, predetermined values are changed and the whole process starts over.

Parameters subject to optimization procedure are interim air gap thickness D3, aperture slot width $Y_y$, aperture slot length $X_y$, phasing stub width $w_s$, matching stub width $W_l$, and  matching stub length $L_l$~(Fig.\ref{Fig:unit_cell}). Optimization procedure is conducted for these six parameters. Results and discussions are given in next section.

\begin{table}
\renewcommand\arraystretch{1.3}
\caption{Predetermined Design Parameters}
\begin{center}
\begin{tabular}{c|c}
\hline
Parameters & Values \\
\hline
\hline
Operating frequency/wavelength & 10 GHz / 30 mm\\
\hline
Scaling factor $S$ & 0.2\\
\hline
Substrate & Rogers RO4003C™\\
\hline
Dielectric constant $\epsilon_r$ & 3.38 \\
\hline
Dissipation factor tan$\delta$ & 0.0027\\
\hline
Substrate thickness D2 and D4&0.508~mm\\
\hline
Cell size&0.4$\lambda \times$0.4$\lambda$\\
\hline
D1 air gap & $\lambda /$4\\
\cline{1-2}
\end{tabular}\label{tab:predetermined}
\end{center}
\vspace{-3mm}
\end{table}

\section{Simulation Results for Unit Cell}

\begin{figure}[!hbtp]
\centerline{\includegraphics[width=0.7\columnwidth]{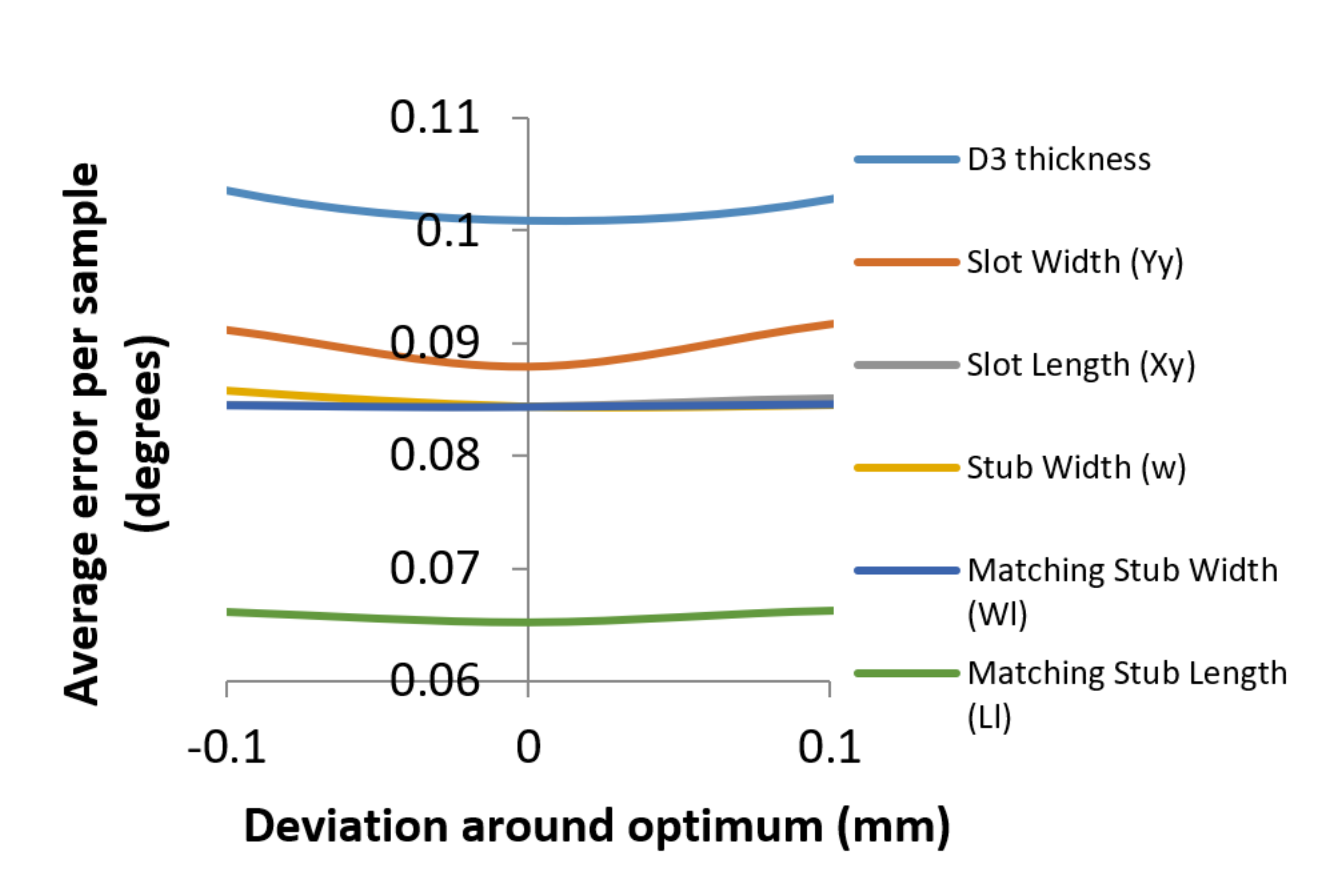}}
\caption{Normalized results for optimization process for all parameters.} 
\label{Fig:simulation_results}
\end{figure}

In the simulations, normal incident wave at 10 GHz is applied with two base linear polarizations. Resonant dimensions are defined as $L$ = 8.0 mm for 0.4$\lambda$ x 0.4$\lambda$ unit cell. For all simulations, $S_{11}$ amplitude is well above -1~dB and phase range is greater than two cycles. Therefore, only phase linearity calculations are shown.

Error rate ($\Delta$) is simulated for all parameters that are subject to optimization with respect to parameter length. They are given in Fig.~\ref{Fig:simulation_results} as normalized values. Values that make the error rate minimum for each parameter are chosen to be optimum. Average error for the last design is 0.065° per sample. After these simulations, first iteration is completed. Optimized lengths are given in Table~\ref{tab:optimized}. 
\begin{table}
\renewcommand{\arraystretch}{1.3}
\caption{Optimized Design Parameter}
\label{tab:optimized}
\centering
\begin{tabular}{c|c}
    \hline
    Parameters & Values~(mm)\\
    \hline
    \hline
    D3 interim air gap thickness &1.6\\
    \hline
    $Y_y$ slot width & 0.9\\
    \hline
    $X_y$ slot length& 10.8\\
    \hline
    $w_s$ stub width & 1.4 \\
    \hline
    $W_l$ matching stub width & 4.0\\
    \hline
    $L_l$ matching stub length &3.0\\
    \hline
\end{tabular}
\end{table}
\begin{figure}[!hbtp]
\centerline{\includegraphics[width=0.7\columnwidth]{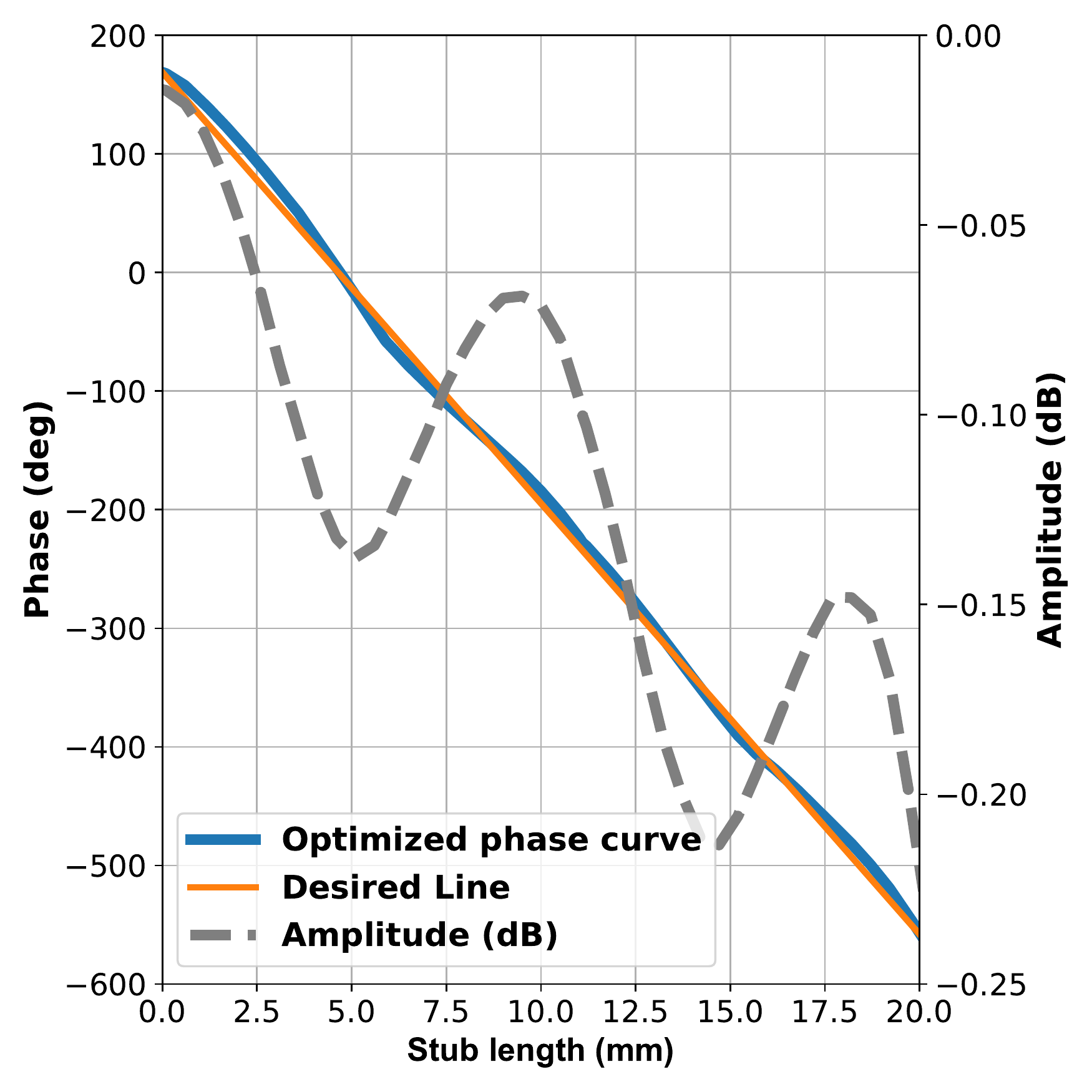}}
\caption{Optimized unit cell $S_{11}$ phase and amplitude.}
\label{Fig:optimized_unit_cell}
\vspace{-3mm}
\end{figure}

Final phase and amplitude curves for unit cell design using optimal values are given in Fig.~\ref{Fig:optimized_unit_cell}. A linear phase curve with more than two cycles range and return loss less than at most 0.25 dB is obtained. Amplitude curve gradually decreases due to ohmic losses and its sinusoidal behavior marks the resonance points. Since linearity and loss level are sufficient, no further iterations are required. Full scale antenna is fabricated using this design.

In Table~\ref{tab:comparison}, the proposed unit cell is compared with those from Refs.~\cite{riel_design_beam_scanning} and~\cite{Carrasco_large_Reflectarrays}. We find that the total phase range is better in Ref. [27]. This is expected since total range increases with the phase stub length which is related to the cell size. On the other hand, using less metal and no circuit element yield a better return loss in our proposed unit cell. 

Fractal configuration results in a smaller patch, thus allows the designer to exploit lowered mutual coupling or a smaller cell size in the teeth of each other. In the end, by this intrinsic miniaturization capability of fractal geometry, a considerably low loss and small size unit cell is successfully designed.
\begin{table}
\renewcommand\arraystretch{1.3}
\caption{Unit Cell Comparison}
\begin{center}
\begin{tabular}{c|c|c|c}
\hline
Parameter & \shortstack{Unit Cell\\ Ref.~\cite{riel_design_beam_scanning}} & \shortstack{Unit Cell\\ Ref.~\cite{Carrasco_large_Reflectarrays}} & \shortstack{Proposed\\ Unit Cell}\\
\hline
\hline
Cell size & 0.42~$\lambda^2$ & 0.28~$\lambda^2$ & 0.16~$\lambda^2$  \\ 
\hline
Avg-Max Loss$^*$&1.3-2.4~dB&0.23-0.32~dB&0.08-0.14~dB
\\
\hline
Total phase range&$373^{\circ}$&$1080^{\circ}$&$740^{\circ}$\\
\hline
Thickness & 0.06~$\lambda$ & 0.34~$\lambda$ &0.34~$\lambda$ \\
\cline{1-4}
\multicolumn{4}{l}{\renewcommand\arraystretch{0.7}\mbox{*}: both average and maximum losses of the works in the table are pro-}\\
\multicolumn{4}{l}{\renewcommand\arraystretch{0.7} vided for the same phase-range of~\cite{riel_design_beam_scanning}.}\\
\end{tabular}\label{tab:comparison}
\end{center}
\vspace{-3mm}
\end{table}

Enhancing the bandwidth is not the primary focus of this study and the gain and efficiency parameters are related to the size of the array as well as the source antenna in use. Thus, comparison with other works is not fully possible. Furthermore, simulated cross polarization value is -45~dB.

\section{Full Size Antenna Design and Analysis}
In order to fully demonstrate the performance of the multilayer Minkowski fractal unit cell with aperture coupled stubs, a 221-element array is designed and fabricated. Phase distribution is calculated with the method explained in~\cite{Ozturk_double_orthogonal}. 

Floquet model assumes that all neighboring cells are identical. Therefore, it is prudent to arrange the stubs such that neighboring cells have closer lengths. One should also choose to reduce the amount of metal parts and keep away from the edges in the structure as much as possible to reduce mutual coupling. 
\begin{figure}[!hbtp]
\centerline{\includegraphics[width=0.7\columnwidth]{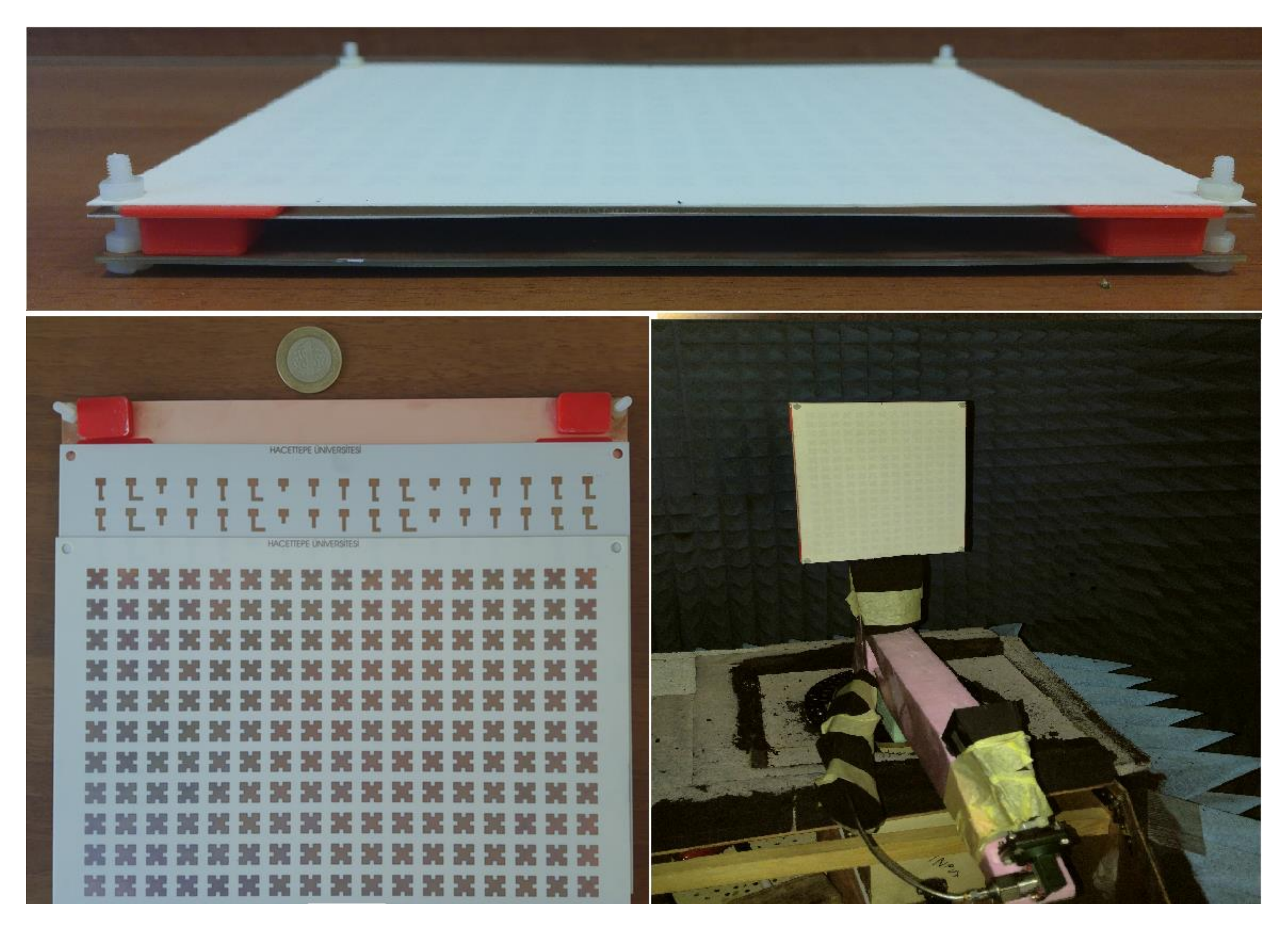}}
\caption{Fabricated reflectarray antenna and measurement setup. }
\label{Fig:photos}
\vspace{-3mm}
\end{figure}

Phase range provided by the unit cell design is around two cycles. First cycle occurs at stub lengths in the first half of its range, (0-10 mm), and the second cycle occurs at stub lengths in the second half, (10-20 mm). Thus, there are two options for all cells when defining the necessary stub lengths. In the first step, all stub lengths are assigned to have lengths in the 0-10 mm range. In this work, we preferred to use stubs that are in the first half of the available range to reduce the error caused by Floquet model which assumes identical neighboring cells and calculates mutual coupling based on this assumption.

Floquet model has given the size of the metal patch ($L$) as 8.00 mm providing the antenna array tuned to the desired frequency. If a uniform square patch is used instead, for the same unit cell size, the dimension would be 9.05 mm which yields a 21.9\% surface size reduction at the cost of 10.12 mm thickness. Fabricated antenna and measurement setup are shown in Fig.~\ref{Fig:photos}.

\section{Experimental Results for The Full Size Antenna}
\begin{figure}[!hbtp]
\centerline{\includegraphics[width=0.7\columnwidth]{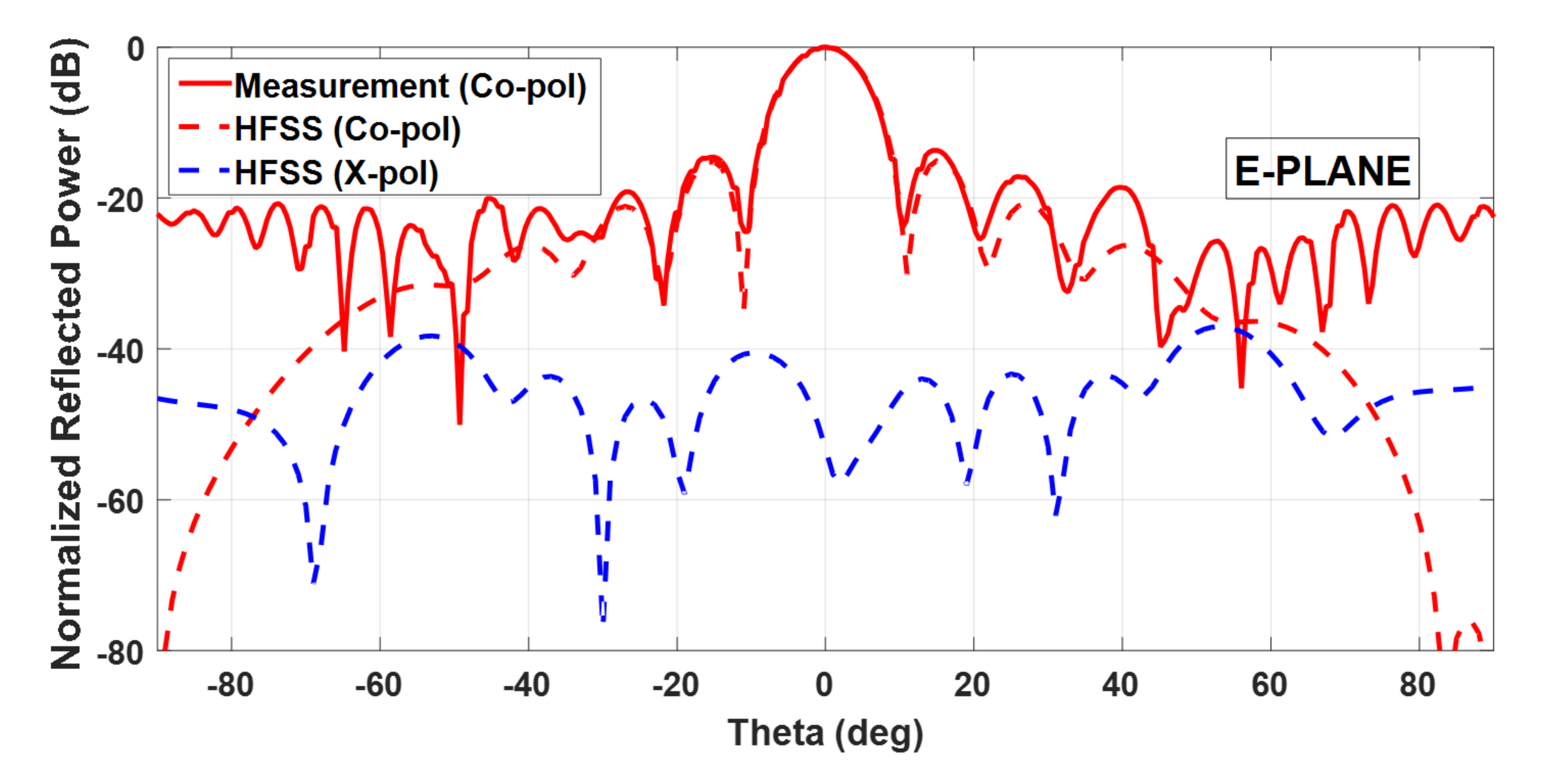}}
\caption{E-plane normalized power pattern. }
\label{Fig:E_plane}
\vspace{-3mm}
\end{figure}
\begin{figure}[!hbtp]
\centerline{\includegraphics[width=0.7\columnwidth]{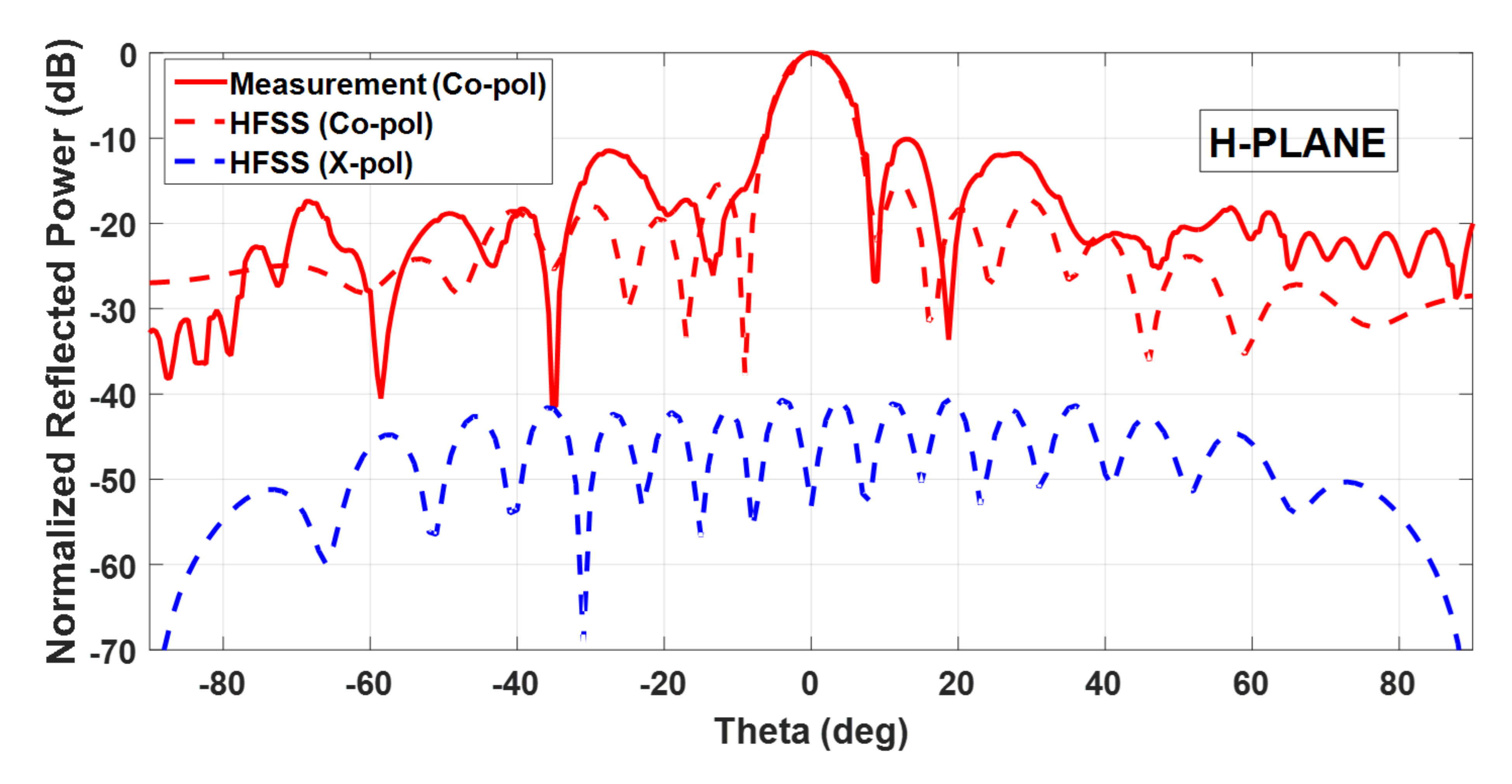}}
\caption{H-plane normalized power pattern.} 
\label{Fig:H_plane}
\vspace{-3mm}
\end{figure}

Normalized co-pol and x-pol power patterns for E and H planes are given in Fig.~\ref{Fig:E_plane} and Fig.~\ref{Fig:H_plane}, respectively. Measured results are in agreement with those obtained from simulations for various parameters (Table~\ref{tab:full_size_params}).

\begin{table}
\renewcommand{\arraystretch}{1.3}
\caption{Full Size Antenna Parameters (E-Plane)}
\label{tab:full_size_params}
\centering
\begin{tabular}{c|c|c}
    \hline
    Parameter & Simulated & Measured\\
    \hline
    \hline
    HPBW&$9.6^{\circ}$&$9.6^{\circ}$\\
    \hline
    SLL & -15.1~dB&-14.2~dB\\
    \hline
\end{tabular}
\end{table}

We have a very good agreement at the main lobe. As the measurement gets closer to horizontal plane, mismatch becomes more visible. On the other hand, distance between two neighboring patches is less than 0.15$\lambda$ and the period of the array is 0.4$\lambda$. Close cell formation has a positive impact on simulated bandwidth (BW), which is measured to be 2.5\%. The main reason for the difference between BW values of the proposed unit cell and the full size antenna is the differential spatial phase delay effect. Since dielectric material thickness is less than 0.02$\lambda$ and no additional technique to enhance the bandwidth is used, we regard this value to be sufficient for our purposes.

Half power beam-width (HPBW) and side lobe level parameters are in very good agreement with the simulations. Cross polarization levels due to the simulation results for both planes are lower than -40~dB and much less than the dynamic range of the anechoic chamber (-20~dB) where the antenna is measured. Therefore, it is of no significance to include measured cross polarization in relevant figures. The incompatibility between co-pol measurements and simulations that are below -20~dB have the same cause. In the H-plane results, feed antenna blockage and spatial reflection effects are observed around the main beam. Since slots are slightly wider than the patch, some of the energy is radiated directly from the slot and not coupled to the patch. This is the main reason for spatial reflection on the H-plane.

Gain and efficiency of a reflectarray antenna depend on factors other than the antenna itself (e.g. the feed antenna). Thus, we choose to give directive gain which represents the performance of only the fabricated antenna. Maximum directive gain is calculated from the measurements as 26.8~dB for the full size reflectarray.

\section{Conclusions}
Using aperture coupled phase stubs is one of the ways of achieving phase range in reflectarray antennas. In this paper, this technique is adapted to Minkowski fractal type of unit cells to overcome the inadequate phase range problem. In the end, phase response wider than two cycles and reflection loss better than 0.25~dB with 21.9\% reduced surface size of the patch is obtained. Reduced surface size of metal patch ensures decreased mutual coupling and better ohmic losses. This yields to reduced error between simulation results using Floquet model and experiments.
Proposed unit cell is used to design and fabricate a 221 element reflectarray antenna. An optimization procedure is identified and implemented to determine the final unit cell design. The full size antenna is simulated using HFSS and compared with measured results. Good agreement is obtained for a wide range of parameters.
\bibliographystyle{IEEEtran}
\bibliography{IEEEabrv, papers}

\begin{thebibliography}{10}
\providecommand{\url}[1]{#1}
\csname url@samestyle\endcsname
\providecommand{\newblock}{\relax}
\providecommand{\bibinfo}[2]{#2}
\providecommand{\BIBentrySTDinterwordspacing}{\spaceskip=0pt\relax}
\providecommand{\BIBentryALTinterwordstretchfactor}{4}
\providecommand{\BIBentryALTinterwordspacing}{\spaceskip=\fontdimen2\font plus
\BIBentryALTinterwordstretchfactor\fontdimen3\font minus
  \fontdimen4\font\relax}
\providecommand{\BIBforeignlanguage}[2]{{%
\expandafter\ifx\csname l@#1\endcsname\relax
\typeout{** WARNING: IEEEtran.bst: No hyphenation pattern has been}%
\typeout{** loaded for the language `#1'. Using the pattern for}%
\typeout{** the default language instead.}%
\else
\language=\csname l@#1\endcsname
\fi
#2}}
\providecommand{\BIBdecl}{\relax}
\BIBdecl

\bibitem{huang2007reflectarray}
\BIBentryALTinterwordspacing
J.~Huang and J.~Encinar, \emph{Reflectarray Antennas}, ser. IEEE Press Series
  on Electromagnetic Wave Theory.\hskip 1em plus 0.5em minus 0.4em\relax Wiley,
  2007. [Online]. Available:
  \url{https://books.google.com/books?id=JNwYCzFt8Z0C}
\BIBentrySTDinterwordspacing

\bibitem{Berry_reflectarray}
D.~{Berry}, R.~{Malech}, and W.~{Kennedy}, ``The reflectarray antenna,''
  \emph{IEEE Trans. on Antennas and Propag.}, vol.~11, no.~6, pp. 645--651,
  1963.

\bibitem{Maligisi1978_microstrip}
C.~S. {Malagisi}, ``{Microstrip disc element reflect array},'' in \emph{EASCON;
  Electron. and Aero. Sys. Conv.}, Jan. 1978, pp. 186--192.

\bibitem{Huang_microstrip_reflectarray}
J.~{Huang}, ``Microstrip reflectarray,'' in \emph{Antennas and Propag. Soc.
  Symp. Digest}, 1991, pp. 612--615 vol.2.

\bibitem{Kelkar_FLAPS}
A.~{Kelkar}, ``{FLAPS}: conformal phased reflecting surfaces,'' in \emph{Proc.
  IEEE National Radar Conf.}, 1991, pp. 58--62.

\bibitem{Pozar_analysis_reflectarray}
D.~M. {Pozar} and T.~A. {Metzler}, ``Analysis of a reflectarray antenna using
  microstrip patches of variable size,'' \emph{Electronics Letters}, vol.~29,
  no.~8, pp. 657--658, 1993.

\bibitem{Huang_Ka_band}
J.~{Huang} and R.~J. {Pogorzelski}, ``A ka-band microstrip reflectarray with
  elements having variable rotation angles,'' \emph{IEEE Trans. on Antennas and
  Propag.}, vol.~46, no.~5, pp. 650--656, 1998.

\bibitem{Huang_bandwidth_study}
J.~{Huang}, ``Bandwidth study of microstrip reflectarray and a novel phased
  reflectarray concept,'' in \emph{IEEE Antennas and Propag. Soc. Int. Symp.
  Digest}, vol.~1, 1995, pp. 582--585 vol.1.

\bibitem{Encinar_design_of_two_layer}
J.~A. {Encinar}, ``Design of two-layer printed reflectarrays using patches of
  variable size,'' \emph{IEEE Trans. on Antennas and Propag.}, vol.~49, no.~10,
  pp. 1403--1410, 2001.

\bibitem{Encinar_three_layer}
J.~A. {Encinar} and J.~A. {Zornoza}, ``Broadband design of three-layer printed
  reflectarrays,'' \emph{IEEE Trans. on Antennas and Propag.}, vol.~51, no.~7,
  pp. 1662--1664, 2003.

\bibitem{Bialkowski_passive_reflectarray}
A.~W. Robinson, M.~E. Bialkowski, and H.~J. Song, ``A passive reflect array
  with dual-feed microstrip patch elements,'' \emph{Microwave and Opt. Tech.
  Letters}, vol.~23, no.~5, pp. 295--299, 1999.

\bibitem{Keller}
M.~G. {Keller}, M.~{Cuhaci}, J.~{Shaker}, A.~{Petosa}, A.~{Ittipiboon}, and
  Y.~M.~M. {Antar}, ``Investigation of novel reflectarray configurations,'' in
  \emph{Symp. on Antenna Tech. and Applied Electromag.}, 2000, pp. 299--302.

\bibitem{riel_design_beam_scanning}
M.~{Riel} and J.~{Laurin}, ``Design of an electronically beam scanning
  reflectarray using aperture-coupled elements,'' \emph{IEEE Trans. on Antennas
  and Propag.}, vol.~55, no.~5, pp. 1260--1266, 2007.

\bibitem{venneri_design_validation}
F.~{Venneri}, S.~{Costanzo}, and G.~{Di Massa}, ``Design and validation of a
  reconfigurable single varactor-tuned reflectarray,'' \emph{IEEE Trans. on
  Antennas and Propag.}, vol.~61, no.~2, pp. 635--645, 2013.

\bibitem{Carrasco_aperture_coupled}
E.~{Carrasco}, M.~{Barba}, and J.~A. {Encinar}, ``Aperture-coupled reflectarray
  element with wide range of phase delay,'' \emph{Electronics Letters},
  vol.~42, no.~12, pp. 667--668, 2006.

\bibitem{Carrasco_slots}
E.~{Carrasco}, M.~{Barba}, and J.~A. {Encinar}, ``Reflectarray element based on
  aperture-coupled patches with slots and lines of variable length,''
  \emph{IEEE Trans. on Antennas and Propag.}, vol.~55, no.~3, pp. 820--825,
  2007.

\bibitem{Borja_sierpinski}
C.~{Borja} and J.~{Romeu}, ``Multiband {S}ierpinski fractal patch antenna,'' in
  \emph{IEEE Antennas and Propag. Soc. Int. Symp.}, vol.~3, 2000, pp.
  1708--1711 vol.3.

\bibitem{Romeu_Koch}
J.~{Romeu}, C.~{Borja}, and S.~{Blanch}, ``High directivity modes in the {K}och
  island fractal patch antenna,'' in \emph{IEEE Antennas and Propag. Soc. Int.
  Symp.}, vol.~3, 2000, pp. 1696--1699 vol.3.

\bibitem{Zubir_minkowski}
F.~Zubir, M.~Rahim, O.~Ayop, and H.~Majid, ``Design and analysis of microstrip
  reflectarray antenna with {M}inkowski shape radiating element,'' \emph{Prog.
  In Electromag. Research B}, vol.~24, pp. 317--331, 01 2010.

\bibitem{Costanzo_fractal}
S.~{Costanzo} and F.~{Venneri}, ``Fractal shaped reflectarray element for wide
  angle scanning capabilities,'' in \emph{IEEE Antennas and Propag. Soc. Int.
  Symp. (APSURSI)}, 2013, pp. 1554--1555.

\bibitem{Costanzo_miniaturized}
S.~{Costanzo} and F.~{Venneri}, ``Miniaturized fractal reflectarray element
  using fixed-size patch,'' \emph{IEEE Antennas and Wireless Propag. Letters},
  vol.~13, pp. 1437--1440, 2014.

\bibitem{Ozturk_double_orthogonal}
\BIBentryALTinterwordspacing
E.~Ozturk and B.~Saka, ``Double orthogonal phase stubs technique for
  {M}inkowski fractal reflectarray antenna,'' \emph{Journal of Electromag.
  Waves and App.}, vol.~33, no.~5, pp. 601--611, 2019. [Online]. Available:
  \url{https://doi.org/10.1080/09205071.2019.1566027}
\BIBentrySTDinterwordspacing

\bibitem{Wahid_dual_layer_minkowski}
A.~{Wahid}, M.~K.~A. {Rahim}, and F.~{Zubir}, ``Analysis of dual layer unit
  cell with {M}inkowski radiating shape for reflectarray antenna on different
  substrate properties,'' in \emph{IEEE Asia-Pacific Conf. on Applied
  Electromag. (APACE)}, 2010, pp. 1--5.

\bibitem{E_ozturk_phd}
E.~Ozturk, ``Minkowski reflectarray analysis and design at {X}-{B}and,'' Ph.D.
  dissertation, Hacettepe University, 2018.

\bibitem{Pozar_millimeter_wave}
D.~M. {Pozar}, S.~D. {Targonski}, and H.~D. {Syrigos}, ``Design of millimeter
  wave microstrip reflectarrays,'' \emph{IEEE Trans. on Antennas and Propag.},
  vol.~45, no.~2, pp. 287--296, 1997.

\bibitem{Frances_phd}
F.~J. Harackiewicz, ``Electromagnetic radiation and scattering from microstrip
  antennas on anisotropic substrates,'' Ph.D. dissertation, University of
  Massachusetts Amherst, 1990.

\bibitem{Carrasco_large_Reflectarrays}
E.~{Carrasco}, J.~A. {Encinar}, and M.~{Barba}, ``Bandwidth improvement in
  large reflectarrays by using true-time delay,'' \emph{IEEE Trans. on Antennas
  and Propag.}, vol.~56, no.~8, pp. 2496--2503, 2008.

\end{thebibliography}
\end{document}